\documentclass[12pt]{article}
\begin{document}

\begin{centering}
{\LARGE \bf{Ex-nihilo: Obstacles Surrounding Teaching  \\ the Standard Model}} \\
Kevin A.\ Pimbblet \\
Address: Department of Physics, University of Durham, South Road, Durham, DH1 3LE, UK.\\
E-mail: K.A.Pimbblet@durham.ac.uk \\
\end{centering}
  
\section*{Abstract}
  The model of the Big Bang is an integral part of the national curriculum for England.  
  Previous work (e.g. Baxter 1989) has shown that pupils often come into
  education with many and varied prior misconceptions emanating from both internal and
  external sources.  Whilst virtually all of these misconceptions can be remedied, there will
  remain (by its very nature) the obstacle of ex-nihilo, as characterised by the question `how
  do you get something from nothing?'  There are two origins of this
  obstacle: conceptual (i.e. knowledge-based) and cultural (e.g. deeply held religious
  viewpoints).   The article shows how the citizenship
  section of the national curriculum, coming `online' in England from September 2002,
  presents a new opportunity for exploiting these. 
  
  Keywords: High School, Astronomy/Cosmology, Social/Cultural
  
\section{Introduction}
  The model of the Big Bang (The Standard Model herein; e.g.\ Peebles 1993) at Key Stage
  4 is a requisite part of science and physics teaching in England (i.e. contained
  within the framework of the National Curriculum; 3d: `about some ideas used to explain
  the origin and evolution of the Universe').  It is common for trainee teachers to be given
  background information (typically based upon lectures courses and planetarium visits) on
  the teaching of astrophysics and cosmology.  
  
  Even if a teacher is a consummate professional who has been trained
  in the teaching of astrophysics, however, 
  one finds that there is a miasma of obstacles in the way of
  successfully teaching the Standard Model.  Indeed, Comins (1993) prepares a list of over
  550 individual misconceptions related to the teaching of astrophysics.  These
  misconceptions fall roughly into three camps: those caused by incorrect internal (mental)
  processes, those caused by incorrect external processes (inaccurate information from
  teachers, parents and peers), and those caused by a combination of both internal and
  external factors. Whilst there exists plenty of evidence in the literature for the causes of
  such misconceptions, there is little in the way of strategies to deal with them.

\section{Obstacles}
  Research by a number of different authors (e.g. Baxter 1989; Sneider \& Pulos 1983;
  Nussbaum, 1979) suggests that children and juveniles will start to formulate their own
  ideas about the natural world long before any formal education is given.  Inevitably, such
  self generated formulations will be riddled with both internal and external misconceptions.
  As their formal education progresses, these scientific misconceptions should be
  systematically dealt with during the course of time.  For example, Baxter (1989) cites the
  case of a pupil persuaded towards an Aristotelian point of view.
  
  Such misconceptions can readily be corrected.  To this end, Bennett (2000) advocates the
  use of a five step plan based upon the un-learning of these misconceptions.  By
  establishing a contextual framework, Bennett (2000) argues that misconceptions about
  astronomy can expediently be corrected.  
  Consider the case of a pupil harbouring an Aristotelian viewpoint.  Aristotle reasoned that
  the Earth was stationary in space and the Sun, Moon, other planets and stars orbit around
  it in circles.  Later, Ptolomy refined this view with his theory of celestial 
  spheres: the Earth was
  at the centre of the Universe and each extra-terrestrial body moved its own sphere with the
  outermost sphere consisting of all of the (fixed) stars.  Knowing that this left room for
  Heaven and Hell outside of these spheres, the church adopted this theory.
  Observations consistently disagreed with this view, the primary one 
  being retrograde (backwards) movement of planets across the sky.  
  Although attempts were made to understand this movement (e.g. Ptolomy proposed that
  planets may execute epicycles) it was not until the seventeenth century that Galileo,
  having learnt of the invention of and subsequently built his own telescope, 
  provided conclusive observations against the geocentric model of Aristotle.  He showed
  that the satellites of Jupiter (later known as the Galilean satellites) do not
  orbit Earth but orbit Jupiter instead.  Earth need not be at the centre of everything. 
  Indeed, Galileo showed too that Venus exhibits phases, which can be explained
  if both Venus and the Earth orbit the Sun.
  Further, Kepler's development from
  the heliocentric model of Copernicus demonstrated elegantly how retrograde motion
  could be explained in a simple manner.
  Further problematic observations for the geocentric model in the seventeenth 
  century included Galileo demonstrating that sunspots were not shadows of
  planets passing in front of the Sun, rather actual spots on its surface that
  rotated; and the supernova of 1604 whose parallax demonstrated that it was
  located beyond the planets and thus the heavens were subject to change.
  Presentation of these observational arguments would readily
  challenge and correct an Aristotelian viewpoint.  

  There is, however, only so far that this
  technique can be applied before a terminal problem arises.  This obstacle can be
  characterised by the question `how do you get something from nothing?' (the ex-nihilo
  obstacle, if you will).  This question can be divided into two aspects: a conceptual problem
  and a cultural problem.

  Scientific understanding and knowledge of how to tackle the ex-nihilo obstacle is,
  perhaps, at a level above and beyond what an average physics teacher is capable of
  addressing.  Indeed, it is probable that said teacher has had little theoretical grounding in
  cosmology and the Standard Model; depending, of course, where they were trained and
  what courses they undertook as an undergraduate (Baxter 1991; Baxter et al. 1991). 
  Even if they do possess such knowledge, answering the characteristic question given
  above would require great expertise and teaching to effectively give an answer appropriate
  at Key Stage 4.  
  The crucial problem is that the theory of general relativity (e.g. Einstein 1950) requires 
  modification when density tends to infinity (as happens when we reverse time towards the Big Bang; 
  Peebles 1993) to take account of quantum
  effects.  Since we do not possess a complete theory of quantum gravity, we can
  presume that our theories are only valid for times after quantum effects became small.
  This limit of general relativity is the Planck time, about $10^{-43} s$ after the Big
  Bang.
  We can only make extrapolations to the Planckian era based upon theory and
  conjecture, which are typically only accessible at the post-doctoral level.

  Inevitably, by the very nature of cosmology, the answering of such a characteristic
  question is likely to come into conflict with pupils who possess strongly held religious
  beliefs (e.g. creationism).  This conflict cannot be simply resolved by citing scientific
  evidence: these pupils' views amount to belief, and are based on faith.
  
\section{Discussion}
  In attempting to answer the ex-nihilo obstacle, there are several tactics that a teacher may
  employ.  To confront the conceptual problem requires an acknowledgement that scientific
  understanding and knowledge is still being developed to this day.  
  For example, since the Hubble Space Telescope (HST) was launched 1990 it has provided
  astronomers with a plethora of new data to characterize and explain.  Indeed, one of
  the HST key projects were observations of distant supernovae to determine how
  the Universe is expanding.  This led to the startling discovery that the Universe
  is not only expanding, but accelerating (e.g.\ Perlmutter et al. 1999).  
  These observations are forcing
  astronomers to reconsider their cosmological models.  
  We can only guess at what evidence the next generation space telescope 
  may reveal.
  Science is, after all, characterised by evidence-based argument.
  Accordingly, there is no one theory
  that describes everything in the Universe without flaw.  

  Teachers are therefore advised to make appropriate citations to 
  evidence for the Big Bang.  Starting simply, the expansion of the 
  Universe explains why light captured from distant galaxies 
  displays similar line spectra to light in a laboratory but is shifted to longer 
  wavelengths (e.g. Hubble \& Humason 1931). 
  At a more sophisticated level, the cosmological microwave 
  background radiation (CMBR; Penzias \& Wilson 1965)
  provides more evidence for the Big Bang. 

  If the Universe came from a previous epoch that was highly
  compressed and hot, the radiation from that epoch must still
  be with us today.  When the temperature of the Universe
  was billions of Kelvin, such radiation would have been in the
  form of $\gamma$-rays.  As the Universe expanded, however,
  it cooled and the radiation's wavelength increased 
  (recall Wein's law).  Gamow and others calculated that
  this radiation at the present epoch would be in the microwave
  part of the electromagnetic spectrum (e.g.\ if the temperature is
  about 5 K, then the wavelength of the radiation will
  be $\lambda \sim 0.06 cm$).  
  Therefore, interpreted as highly redshifted radiation 
  from a primeval fireball, the ubiquity of the CMBR gives excellent evidence of a previous
  state of the Universe that was extremely hot.
  Both the redshifted Universe and the CMBR can be explained at Key Stage 4 level.
  Whilst the Big Bang model explains these and many other observations, 
  historically it has not been without criticism.  
  For example, one may enquire, why is the Universe so close to the critical density?
  For the moment, scientists accept the Big Bang model
  as it ties together many threads of evidence.  Yet, they must also be sceptical: the
  model has had problems in the past and will no doubt have problems in the future too.
  
  The cultural problem is a much thornier issue.   
  Contained within the declaration of human rights is the right to freedom of religion.  The
  teacher, being in a responsible role, must ensure that they do not force their own
  (scientific) viewpoints upon such a pupil.  Rather, they must allow the pupil to come to
  their own conclusions after presentation of the scientific evidence that supports the
  Standard Model.  To force a viewpoint upon a pupil would be more than counter-productive.  
  Such force could generate resentment towards
  the teacher, and lead to rejection of the subject material without consideration.
  
  Since these dichotic opinions potentially lead to conflict, I advocate
  the use of a full classroom discussion to debate the issues surrounding the ex-nihilo
  obstacle.  A discussion provides excellent opportunities for cross-curricular activity whilst
  allowing the pupils to express their own opinions.  Research skills (to back up any
  scientific argument; for example, see Stephen Hawking's Universe series at www.pbs.org/wnet/hawking), 
  information technology skills (to present ideas with) and literacy can
  all play a significant role in such a discussion.  Other subject areas could readily be drawn
  upon as well (e.g.\ drama).  
  
  More significantly, with the subject of citizenship coming `online' in England from September 2002,
  an already squashed timetable could readily accommodate aspects of citizenship within an
  astronomical origins debate.  Particular items in the citizenship national
  curriculum programme of study include (at Key Stage 4) 1b (`the origins and
  implications of the diverse national, regional, religious and ethnic identities in the United
  Kingdom and the need for mutual respect and understanding'); 2 (all points under
  `developing skills of enquiry and communication' relevant).  Not only will
  such a discussion enhance pupil development, it would also actively promote citizenship
  though astronomy.  It must be emphasized, however, that pupils need to be carefully guided
  to preserve mutual respect.
  
\section{Conclusions}
  This work has reviewed some research on the misconceptions encountered during the
  teaching of astrophysics.  When corrected, these misconceptions will generally be dealt
  with effectively, but there will often remain an ex-nihilo obstacle.
  
  Such an obstacle can be divided into two separate issues: a conceptual (knowledge-based)
  one and a cultural one.  In tackling these issues, one must address:\\ 
\ \\
  1. Science is still developing to this day and we do not presently possess a theory that
       describes everything without flaw.\\
  2. Some pupils will hold religious values that may appear contradictory to the Standard
       Model.\\
\ \\  
  Since both of these points lead to a potential conflict and debate, a full 
  discussion is therefore advocated, simultaneously developing aspects of the 
  national curriculum for citizenship.
  
\section*{Acknowledgements}
  The preparation of this manuscript has made use of resources at both the University of
  Durham and St.Leonard's R.C.V.A. School, Durham.  KAP thanks Mary C.\ Hawkrigg for
  her reading through of the text and for providing useful suggestions that have improved
  this work.  Finally, thanks to John Newman, Marion E.\ Jones, Peter Little and 
  Simon B.\ Campbell for
  most unknowingly inspiring this work and for providing copious amounts of feedback on
  my teaching.
  
\section*{References}
  
  Baxter, J., 1989, International Journal of Science Education, 11, 502 \\
  Baxter, J., 1991, Quart. Journal of the Royal Astron. Soc., 32, 147 \\
  Baxter, J.,~et al., 1991, Quart. Journal of the Royal Astron. Soc., 32, 159\\
  Bennett, J., 2000, American Astron. Soc. Meeting, 197, 86.01\\
  Comins, N.F., 1993, Bulletin of the American Astron. Soc., 25, 1430\\
  Einstein, A., 1950, `The Principle of Relativity', Metheun, London\\
  Hubble, E.~\& Humason, M.~L.\ 1931, Astrophysical Journal, 74, 43 \\
  Nussbaum, J., 1979, Science Education, 63, 83\\
  Peebles, P.J.E., 1993, `Principals of Physical Cosmology', Princeton Series in Physics, Princeton University Press, NJ\\
  Penzias, A.~A.~\& Wilson, R.~W.\ 1965, Astrophysical Journal, 142, 419 \\
  Perlmutter, S.~et al., 1999, Astrophysical Journal, 517, 565 \\
  Sneider, C. \& Pulos, S., 1983, Science Education, 67, 205\\

\end{document}